\documentclass[graybox,natbib,nosecnum]{svmult}
\bibpunct{(}{)}{;}{a}{}{,} % suppress commas between author-names and year
\pdfoutput=1 
\usepackage{mathptmx}       % selects Times Roman as basic font
\usepackage{helvet}         % selects Helvetica as sans-serif font
\usepackage{courier}        % selects Courier as typewriter font
\usepackage{type1cm}        % activate if the above 3 fonts are
\usepackage{makeidx}         % allows index generation
\usepackage{graphicx}        % standard LaTeX graphics tool
\usepackage{multicol}        % used for the two-column index
\usepackage[bottom]{footmisc}% places footnotes at page bottom
\usepackage[normalem]{ulem}	% for strike-through of text with \sout{}  
\usepackage{hyperref}  %for hyperlinks
\usepackage{bm}
\usepackage{xspace}
\usepackage{lineno}

\usepackage{soul}   % for high-lighting of text

\usepackage{booktabs}
\usepackage{amsmath}

\newcommand{\hbindex}[1]{\hl{#1}\index{#1}}  %highlights index entries

\newcommand{\lnlike}{log-likelihood\xspace}

\newcommand{\pvec}{\ensuremath{\bm{\theta}}\xspace}
\newcommand{\dpnt}{\ensuremath{y}\xspace}
\newcommand{\dvec}{\ensuremath{\bm{y}}\xspace}
\newcommand{\dmat}{\ensuremath{\bm{Y}}\xspace}
\newcommand{\xvec}{\ensuremath{\bm{x}}\xspace}
\newcommand{\xmat}{\ensuremath{\bm{X}}\xspace}
\newcommand{\ud}{\ensuremath{\mathrm{d}}}
\newcommand{\pr}{\ensuremath{\mathrm{Pr}}}
\newcommand{\covmat}{\ensuremath{\bm{\Sigma}\xspace}}

\makeindex             % used for the subject index
\begin{document}
\title*{Bayesian Methods for Exoplanet Science}
\author{Hannu Parviainen}
\institute{Hannu Parviainen \at Instituto de Astrof\'\i sica de Canarias, C. Via Lactea S/N, E-38205, La Laguna, Tenerife, Spain;  \and 
 Dpto. de Astrof\'isica, Universidad de La Laguna, 38206,La Laguna, Tenerife, Spain; \email{hannu@iac.es}}
\maketitle

\abstract{
	Exoplanet research is carried out at the limits of the capabilities of current telescopes and instruments.
	The studied signals are weak, and often embedded in complex systematics from instrumental, telluric, and 
	astrophysical sources. Combining repeated observations of periodic events, simultaneous observations with 
	multiple telescopes, different observation techniques, and existing information from theory and prior research 
	can help to disentangle the systematics from the planetary signals, and offers synergistic advantages over
	analysing observations separately.
	Bayesian inference provides a self-consistent statistical framework that addresses both
	the necessity for complex systematics models, and the need to combine prior information and heterogeneous 
	observations. 
	This chapter offers a brief introduction to Bayesian inference in the context of exoplanet research, with focus
	on time series analysis, and finishes with an overview of a set of freely available programming libraries.
}

\section{Introduction}
\label{sec:introduction}

Statistical inference has a major role in the interpretation of astrophysical observations. 
%In model-fitting problems
%we have a physical model (that we believe is sufficient to explain the observations), and we aim to infer the model
%parameters from observational data. 
Exoplanet research is generally carried out
at the limits of the capabilities of current telescopes and instruments, and the planetary signals of interest are 
weak and embedded in complex systematics (noise) from instrumental, telluric, and astrophysical sources. 
The reliability of the information 
inferred from the observations depends on how well we understand the statistical characteristics of the observations, 
and on how well these characteristics are taken into account by the methods used to carry out the inference. 

Of the two major schools of statistical inference, frequentist and Bayesian, the latter has gained popularity in physics 
over the last decades. The reasons for the increasing interest in Bayesian methods are manifold: first, Bayesian inference 
offers a consistent approach for combining observational information from different types of observations (e.g., radial
velocities, ground- and space-based photometric time series, etc.) with prior information; second, Bayesian inference allows 
for versatile modelling of the observational uncertainties (errors), facilitating robust analyses; and third, Bayesian inference 
offers an unified, self-consistent, approach for parameter estimation and model comparison. However, Bayesian inference is 
in general computationally more demanding than the frequentist approaches, and its wide adoption has been made possible only 
by the recent advances in numerical methods and computing power.

The Bayesian approach assigns a probability to a hypothesis. A prior probability distribution is used to encode our prior 
information about the hypothesis, and this prior is updated using observations to obtain a posterior probability distribution.
When new data is acquired, the posterior based on the previous data can be used as a prior distribution, and the new data 
is used to obtain a new, updated, posterior distribution.

The problems in statistical inference can be roughly divided into parameter estimation and model comparison. Bayesian parameter 
estimation aims to infer the joint posterior distribution for model parameters given a model (that we believe is sufficient to explain the observations), 
prior information, and observations, while model comparison aims to find which of the possible models is best-suited to explain the observations.

\hbindex{Bayesian parameter estimation} results in a joint posterior distribution for the model parameters. This offers improved versatility
over frequentist methods that generally yield point estimates, since the ways to summarise the posterior can be tailored to take its
complexity into account. For example, an approximately normal posterior can be summarised by its mean and central 68\% posterior interval
to yield the common mean and $1\sigma$ uncertainty estimate; while a powerlaw-like posterior can be summarised by fitting an appropriate
analytic distribution; and a more complex posterior can be described by a set of percentiles, or as a mixture of analytic distributions.
A common case in exoplanet characterisation is to encounter parameters (eccentricity, impact parameter, etc.) where the posterior is 
close-to constant for small parameter values, and starts tapering off towards large values after some point. In these cases the data can
only constrain the upper boundary for the parameter, and reporting a point estimate (other than the upper boundary corresponding to 
some posterior percentile) would make little sense.

In \hbindex{Bayesian model comparison} setting we are usually either interested in finding out how many physical signals a dataset contains (e.g., radial
velocity planet searches), or what is the most likely physical cause for a signal (statistical validation of planet candidates found by transit and RV surveys). Other
use cases exists, but these two are currently the dominant uses for model comparison in exoplanet research.
In the first case, we want to find out when to stop introducing complexity (hypothetical planets) to the model. 
We are not only interested in finding the model that best fits the given data, but we want to find the model 
that explains the actual information content in the data without overfitting (fitting the noise). In the latter case, the models
can be of similar complexity (e.g., a planet candidate signal can be caused by a bona-fide planet or a blended eclipsing binary), 
but the model comparison needs to combine evidence from different data sources. For example, the probability that 
a transiting planet candidate is a real planet can be estimated by combining information from the discovery light curve (transit 
shape, secondary eclipse depth, phase variations, etc.), Milky Way population synthesis simulations (blending probability given 
the aperture size and host star location), stellar characterisation, ground-based follow-up observations, etc.

This chapter aims to offer an introduction to the use of Bayesian inference in exoplanet research, 
with a focus on time series analysis (photometry and radial velocity observations). We present the basic concepts,
theory, and methods, and also overview some issues specific to exoplanet research. However, given the depth of the
topic, the chapter can only scratch the surface of Bayesian inference, and does not discuss the more advanced topics. 
Instead, we aim to direct the reader to more in-depth books and publications thorough the chapter, and finish with an 
overview to useful literacy.

\section{Theory}
\label{sec:theory}
The goal of \hbindex{Bayesian inference} is to make probability statements about unobserved quantities 
conditional on observed data \citep[freely citing][Sect.~1.3]{Gelman2013}. In the case of parameter
estimation (model fitting), we want to infer the joint probability distribution for the model parameters 
given observational data and prior information. This probability 
distribution is called the \textit{posterior distribution} (posterior), and it is obtained by 
updating a \textit{prior distribution} (prior) with a \textit{sampling distribution} (also known as the 
\textit{likelihood}, or \textit{data distribution}.) 

We follow the notation laid out by \citet[][Sect.~1.2]{Gelman2013}, where \pvec is a vector
containing the unobservable quantities (model parameters), \dvec is a vector containing 
the observed data (e.g., RV measurements or photometry), and \xmat is a matrix containing
the explanatory variables (also known as covariates, independent variables, or predictors) 
for the observations. Both the model parameters and observations are modelled as random
quantities, but the covariates are considered to be known exactly. In general, the elements of \dvec 
can be vectors themselves,
in which case we have a matrix of observations \dmat. This is, however, not that usual
in exoplanet research (transmission spectroscopy being an exception, where the observed dataset
consists of a set of narrow-band light curves constructed from a spectroscopic time series), so
we consider each \dvec element as scalar for simplicity. The covariates--such as the mid-exposure
time, airmass, and seeing, when dealing with ground-based observations--are stored as vectors in covariate matrix, 
\xmat, one per observation. However, if we have only a single covariate (usually time in time series 
analysis), we represent the covariate vector as \xvec.

\subsection{Posterior distribution}
\label{sec:theory.posterior}

The \hbindex{posterior distribution} encodes the information about the model parameters given the prior information and the likelihood from observations.
Posterior probabilities for individual parameters are obtained through marginalisation, where the posterior is integrated over all other
parameters than the parameter of interest. Finally, different summary statistics can be derived to describe the posterior distributions
concisely.

The joint posterior distribution can be derived starting from the \hbindex{Bayes' theorem}
\begin{equation}
\pr(H|D) = \frac{\pr(H) \pr(D|H)}{\pr(D)},
\end{equation}
where $\pr(H|D)$ is the posterior probability for a hypothesis $H$ given data $D$, $\pr(H)$ is the prior probability for the hypothesis, 
$\pr(D|H)$ is the probability for the data given the hypothesis, and $\pr(D)$ is the probability for the data. In parameter estimation setting,
the hypothesis $H$ is a (continuous) model parameter vector \pvec, the probabilities are  continuous probability distributions, and the 
posterior distribution is
\begin{equation}
P(\pvec | \dvec) = \frac{P(\pvec) P(\dvec | \pvec)}{P(\dvec)} =  \frac{P(\pvec) P(\dvec | \pvec)}{\int P(\pvec) P(\dvec) \ud \pvec}.
\label{eq:bayes_theorem}
\end{equation}
The integral in the denominator, probability for the data, is  a normalising constant called marginal probability (or model evidence), 
and ensures that the posterior integrates to unity.

\subsection{Prior distribution}
\label{sec:theory.prior}

The role of a \hbindex{prior distribution} is to encapsulate the current information and assumptions about a model parameter
(or a quantity that depends on the model parameters). 
As new information (observations) is obtained, the prior is updated by the likelihood to produce a posterior distribution, 
which can be used as a prior distribution in future analyses. 

Priors can be (roughly) classified as either \hbindex{informative priors} or \hbindex{weakly informative (uninformative) priors}, depending on how strongly they constrain the 
parameter space. Informative priors can be based on previous research and theory. For example, one can use a normal distribution with mean and 
standard deviation based on previously reported parameter mean and uncertainty estimates.  Weakly informative priors are used to express our 
ignorance about a parameter, and aim to minimise the effect the prior has on the posterior, allowing the data to 'speak for itself'. 

The choice of a prior distribution is not objective. When setting informative priors based on previous research, one needs to decide 
how much the reported values can be trusted. For example, if several competing estimates exists, it may be better to create a prior that encompasses 
all the estimates, rather than base a prior on any single estimate. In any case, it is important to report the priors used in an analysis, 
and, unless trivial, describe the justification for the priors (i.e., how and why the priors were chosen). It is also a good practice to test 
how sensitive the analysis is on the priors. Sensitivity analysis is especially important in a model selection setting, where the priors generally 
have a larger impact than in parameter estimation.

\subsection{Likelihood}
\label{sec:theory.likelihood}

The \hbindex{likelihood} represents the probability that the observations follow from a given model evaluated at 
a given point in the model parameter space.
In parameter estimation setting, the model generally consists of a deterministic and a stochastic part. The deterministic part models the 
signals that can be modelled using a parametric model, and the stochastic part aims to explain the noise, 
that is, everything not explained by the parametric model. 

\subsubsection{White noise}
\label{sec:theory.likelihood.white_noise}

If the observations can be explained by a parametric model with additive \hbindex{uncorrelated (white) noise}, 
the joint likelihood is a product of independent likelihoods,
\begin{equation}
 P(\vec{d} | \pvec) = \prod_i P(d_i| \pvec), \label{eq:wn_likelihood}
\end{equation}
where $\vec{d}$ is the dataset, $d_i$ are the individual observations and \pvec is the model 
parameter vector.  The product in Eq.~\ref{eq:wn_likelihood} can easily lead to 
numerical under- or overflows, and it is common to work with log densities instead,
so that
\begin{equation}
 \ln P(\vec{d} | \pvec) = \sum_i \ln P(d_i| \pvec). \label{eq:wn_lnlikelihood}
\end{equation}

As an example, if the noise is normally distributed, a single observation $\dpnt_i$ follows
\begin{equation}
\dpnt_i \sim \mathcal{N}(m(\pvec,\xvec_i), \sigma_i),
\end{equation}
where $\mathcal{N}$ stands for the normal distribution, $m$ is the parametric model (in this case also called the mean function), \pvec 
are the model parameters, \xvec is the vector of covariates for observation $i$, and
$\sigma_i$ is the standard deviation (uncertainty) for observation $i$. The likelihood 
of a single observation is now
\begin{equation}
P(\dpnt_i|\pvec, \xvec_i, \sigma_i) = \frac{1}{{\sigma_i \sqrt{2\pi}}} \exp -\frac{\left(\dpnt_i -m(\pvec, \xvec_i)\right)^2 }{2\sigma_i^2},
\end{equation}
and the log likelihood of a dataset consisting of $n$ observations is
\begin{equation}
  \ln P(\dvec|\pvec,\xmat, \bm{\sigma}) = -\frac{1}{2}\left(n\ln2\pi +\sum_i^n \ln 
\sigma_{\mathrm{i}}^2 + \sum_{i=1}^n \frac{r_i^2}{2\sigma_{\mathrm{i}}^2} \right ), \label{eq:wn_joint_white_likelihood}
\end{equation}
where $r_i$ is the residual, $r_i = \dpnt_i - m(\pvec, \xvec_i)$. 

\subsubsection{Correlated noise}

Unfortunately, observational noise is rarely white. The instrument, Earth's atmosphere, and
different astrophysical processes all give rise to signals, of which only a part can
be represented as simple functions of quantities measured simultaneously with the observations (covariates).
The rest of these noise signals aggregate into time-correlated (red) noise that has to be accounted for
in the analysis using statistical methods. If not accounted for, \hbindex{red noise} can lead to biased parameter
estimates with underestimated uncertainties, and false detections in planet searches. For example, 
stellar granulation can lead to time-correlated photometric variability with amplitude and time-scale 
comparable to planetary transits, while star spots give rise to RV signals that can be mistaken as of 
planetary origin.

\hbindex{Correlated noise} can be represented as a stochastic (random) process in time, and if the noise process 
follows a normal distribution, it can generally be modelled as a \hbindex{Gaussian process} 
\citep[GP,][]{Rasmussen2006,Roberts2013,Gibson2011a}. Generic Gaussian processes require an 
inversion of an $n \times n$ covariance matrix, where $n$ is the number of time series datapoints, 
which does not scale well for large time series. However, methods have been developed to evaluate
temporally correlated GPs with better scaling, which allow GPs to be used in most time series analysis
problems encountered in exoplanet research.

\runinhead{Generic Gaussian Processes}
The scalar \lnlike Eq.~\ref{eq:wn_joint_white_likelihood} can be written in a more general vector form as
 \begin{equation}
  \ln P(\dvec|\pvec) = -\frac{1}{2} \left(n \ln 2\pi +\ln|\covmat| +\vec{r}^\mathrm{T} \covmat^{-1}  \vec{r}\right),
  \label{eq:lnlikelihood_gn}
 \end{equation}
where $n$ is the number of datapoints, $\vec{r}$ is the residual vector, and $\covmat$ is the covariance
matrix. The covariance matrix is diagonal for white noise, which yields Eq.~\ref{eq:wn_joint_white_likelihood} as
a special case, but contains off-diagonal elements when the noise is correlated.

Gaussian processes (GPs) offer a versatile way to model normally distributed 
stochastic processes, with extensions existing for Student-t-distributed processes \citep{Shah2014}. In GP formalism, the
%Input covariates...
covariance matrix elements are given by 
\begin{equation}
\covmat_{i,j} = k(\vec{x}_i,\vec{x}_j) + \sigma^2 \delta_{ij},
\end{equation}
where $k$ is called the covariance function (or kernel), \vec{x} are input parameter vectors, $\delta$ is the Kronecker
delta function, and $\sigma^2$ is the white noise term for the $i$th datapoint. The covariance function maps the input
vectors to a scalar covariance, and thus defines the behaviour of the GP.

Gaussian processes are versatile: they can be used with any number of inputs and outputs; they can be used to model processes 
that combine aperiodic, periodic and quasi-periodic behaviour; and they can be used to extract different additive processes
from a time series. However, in its most general form, a GP evaluation requires the inversion of a covariance matrix, which
has a time complexity $\mathcal{O}(n^3)$, where $n$ is the number of input points. 

Despite the bad scaling properties, GPs have been used in exoplanet research extensively. For example, \cite{Gibson2011a} introduced Gaussian 
processes to the exoplanet research community as a robust tool to model systematics in transmission spectroscopy (with an excellent 
overview of the GP basics in the Appendix); \citep{Rajpaul2015} demonstrated how GPs can be used to disentangle the stellar activity
signal from planetary signal(s) in radial velocity time series; and \citep{Czekala2017} showed how GPs can be used to model stellar 
spectra to improve radial velocity measurements.  

The $\mathcal{O}(n^3)$ scaling restricts the use of generic GPs to problems with a relatively small number of datapoints.
However, the matrix inversion can be accelerated (or bypassed completely) for several special cases, which are discussed
below.

\runinhead{Temporally correlated noise with power law power spectral density}
\citet{Carter2009} introduce a wavelet-based method to calculate the likelihood for a time series having temporally correlated noise 
with power spectral density (PSD) varying as $1/f^\gamma$. The approach uses fast wavelet transform (FWT) and computes the likelihood
in linear time (that is, the method has time complexity $\mathcal{O}(n)$). The method is implemented in many transit modelling packages,
and offers a good alternative to the computation-heavy GPs if the time sampling is approximately uniform, time is used as the only input 
parameter, and if the noise PSD follows a power law. However, more versatile methods should be used if any of these restrictions
is not met.

\runinhead{Temporally correlated noise with a (nearly) arbitrary kernel}
The computational cost of Gaussian processes can be alleviated in the special case when time is the only input parameter, and the
kernel yields covariance matrices satisfying special conditions. 
\citep{Ambikasaran2014} describe an approach that allows for the covariance matrix inversion in $\mathcal{O}(n \log^2 n)$ time,
and covariance matrix determinant computation in $\mathcal{O}(n \log n)$  time. This difference in scalability allows the approach to be used in the
analysis of large time series, such as with Kepler light curves.

\runinhead{Temporally correlated noise with kernel consisting of complex exponentials}
Describing the covariance function as a mixture of complex exponentials allows one-dimensional GPs to be evaluated in linear 
time \citep{Foreman-Mackey2017}. 
Despite its restrictions, such a kernel is flexible enough to model a wide range of astrophysical
variability, and is especially well-suited to model quasiperiodic behaviour. 

\subsection{Marginal likelihood}
\label{sec:theory.evidence}

\hbindex{\textit{Marginal likelihood}}, also known as \hbindex{\textit{model evidence}}, can be ignored in parameter estimation setting, but
is an important quantity in Bayesian model comparison. The marginal likelihood, $Z$, is obtained by integrating the posterior
over the whole parameter space
\begin{equation}
Z = \int P(\pvec)\;P(\dvec|\pvec)\;\ud\pvec.
\end{equation}
The integration is rarely analytically tractable, and generally requires specialised numerical methods.

\section{Parameter estimation}

The aim of \hbindex{model parameter estimation} is to obtain an estimate of the joint posterior distribution for the 
model parameters given the data and prior information.  Per-parameter marginal posteriors can then be 
derived from the joint posterior, and reported concisely using different summarisation methods.

\subsection{Marginal posterior distribution}

The marginal likelihood can be ignored in the parameter estimation
setting, allowing us to work with  an unnormalised joint posterior density
\begin{equation}
P(\pvec | \dvec) = P(\pvec) P(\dvec | \pvec).
\end{equation}
The posterior distribution for a single parameter $\theta_i$ is obtained by integrating the 
joint posterior density over all other parameters
\begin{equation}
P(\theta_i |\dvec) = \int P(\pvec|\dvec) \;\ud\pvec_{j \neq i}.
\end{equation}
This is called marginalisation, and results in a \hbindex{\textit{marginal posterior distribution}} for 
the parameter.

\subsection{Summarising posteriors}
\label{sec:theory.summarising_posteriors}

Bayesian inference is generally based on a random sample drawn from the posterior, which can then be
described concisely using summary statistics. Common point estimates, such as the mean and the mode 
(or modes, if the posterior is multimodal), and a set of posterior percentiles all convey useful information.
Especially, \citet[][see Sects.~2.3 and 10.5]{Gelman2013} recommend reporting the 2.5\%, 25\%, 50\%, 
75\%, and 97.5\% percentiles, which provide the posterior median, 50\% and 95\% posterior intervals, and 
also information about the posterior skewness.

If a marginal posterior can be approximately modelled by an analytical distribution, then the parameters of 
the distribution fitted to the samples can be reported. However, even then, it is useful to report the posterior 
percentiles to allow for easy comparison with other studies.

When practical, it is useful to visualise the marginal posteriors and joint pairwise posteriors. 
This can be done directly from the samples using either 1D and 2D histograms or kernel density estimation 
(KDE), and provides insight into the parameter distributions, and whether there were any
problems in the sampling process.

\subsection{Estimating posteriors: Markov chain Monte Carlo}
\label{sec:parameter_estimation.mcmc}

\hbindex{Markov chain Monte Carlo} (MCMC) sampling is a fundamental tool for posterior estimation when we cannot
sample from the posterior directly. The method produces a set of samples drawn from the posterior by iteratively
constructing a Markov chain with the posterior distribution as its equilibrium distribution.

The sampler starts a chain from a point in the parameter space, proposes a move to another point, 
and either accepts or rejects the move based on the posterior density ratios between the current and proposed 
locations. The location after the proposal is added to the chain (no matter whether the proposal was accepted or 
rejected), and the sampling continues by proposing a new step. The proposal move is constructed in a way to 
satisfy a set of conditions that ensure that the distribution of the samples in the chain asymptotically converges 
to the posterior distribution, but the speed of convergence depends on the complexity of the posterior and the 
MCMC implementation (see \citealt[][Sect.~11]{Gelman2013}; \citealt[][Sect.~6.3]{Robert2007}; and
 \citealt[][Sect.~12]{Gregory2005} for a comprehensive explanation; and \citealt{Betancourt2017} for a
 historical overview).

\runinhead{General considerations}
Markov chain Monte Carlo is, actually, an umbrella term for a family of sampling methods. The basic MCMC
approaches, such as the Metropolis-Hastings algorithm \citep{Hastings1970}, are easy to code,
but the more advanced methods, such as NUTS  \citep{Hoffman2011}, are not straightforward to 
implement by oneself. MCMC sampling packages exist for all the major programming environments, some of which 
are listed in Sect.~\ref{sec:tools}, and usually the first step to do when starting an analysis is to choose
the best sampler applicable to the problem at hand. A good sampler can explore the posterior space efficiently, 
with a minimal number of posterior evaluations. The modern MCMC samplers, like NUTS,
can be very efficient if the posterior derivatives can be calculated analytically. However, if this is not possible,
a less-advanced sampler needs to be used.

While the different sampling approaches have their own peculiarities, some basic steps  for obtaining a reliable
posterior sample can be considered common:
\begin{enumerate}
	\item Try to set up the problem in a way that allows for efficient sampling. For example, choose a sampling 
		parametrisation that reduces mutual correlations between parameters, as discussed in Sect.~\ref{sec:practical_issues.parametrisation}.
		Bad model parametrisation can make it impossible for a sampler to explore the whole posterior space.
	\item Test the sampler by simulating a set of short chains and check the acceptance rate and parameter 
		autocorrelation lengths. See if the sampler has parameters that can be tuned to improve the sampling, or 
		try to reparametrise the problem if the sampler is not able to explore the posterior properly. Modern MCMC
		packages also often offer an automatic tuning functionality, which aims to optimise the sampling.	
	\item Run multiple chains starting from different locations in the parameter space. Simulating multiple chains 
		allows one to test for chain convergence, and can reveal posterior multimodality.
	\item Test for chain convergence using numerical convergence tests. Most MCMC packages implement several
		convergence tests. If the chains have not converged, continue sampling.
	\item Inspect the chains and parameter posteriors visually. This may not always be practical, e.g., 
		if the MCMC simulations need to be repeated for many targets, but visual inspection may give useful
		insight about the convergence, sampling, and warm-up phase.
\end{enumerate}

\runinhead{Acceptance rate}
The fraction of accepted jump proposals to the number of total proposals (\hbindex{acceptance rate}) is an useful 
diagnostic of the sampler's efficiency. For multidimensional models, the acceptance rate should be close 
to 23\% \citep{Gelman1996,Roberts1997}.

\runinhead{Chain warm-up}
With a finite chain, the starting location can affect the chain's convergence. For example,
if starting the chain far from the posterior mode, it may take time before the chain starts sampling the
high-probability posterior space. The influence of the starting point is reduced by excluding an initial
\hbindex{warm-up (or burn-in) period} \citep[][p.~282]{Gelman2013}. There are no hard rules for deciding the
appropriate length for the warm-up period, but it generally can be decided based on convergence tests
and visual inspection of chains.

\runinhead{Autocorrelation and thinning}
The consecutive samples in the MCMC chain are not independent, but correlated, and the strength of correlation
depends on the sampler's ability to sample the posterior.
A set of independent samples can be obtained by \hbindex{\textit{thinning}} the chain, i.e., by selecting every $n$th sample,
where $n$ is close to the chain autocorrelation length. However, chain thinning is not strictly necessary if the 
chains have converged \citep[][p.~282]{Link2012,Gelman2013}, but can still be practical for memory and disk-space saving purposes. 

\runinhead{Testing chain convergence}
The MCMC chain approximates the posterior asymptotically. However, the number of steps needed to obtain a
representative sample depends on how efficiently the sampler can cover the posterior, and the chains need
to be tested for \hbindex{convergence} before they can be used for inference. Converged chains should be statistically
similar to each other, and thus a set of chains started from different points in the parameter space allows us to
test for convergence of the chains in the set. The tests can be further improved by splitting the individual chains 
into two parts (after removing the warm-up phase), since the halves should be similar to each other if the chains
are stationary and well-mixed.

The Gelman-Rubin diagnostic \citep[][also, \citealt{Gelman2013}, p.~284]{Gelman1992} offers a practical 
way to test convergence using a set of chains. The diagnostic compares the estimate for the marginal posterior 
variance $V$ for parameter $\theta$ to the 
mean within-chain variance $W$. The two should be approximately equal for a set of well-converged chains, 
and the \textit{estimated scale reduction} for $M$ chains with $N$ steps,
\begin{equation}
	\sqrt{\hat{R}} = \sqrt{\frac{N-1}{N} + \frac{M+1}{NM} \frac{B}{W}},
\end{equation}
should be close to unity (we have dropped the 
factor $\mathrm{df}/(\mathrm{df}-2)$ from \citealt[][Eq.~20]{Gelman1992}). Here $B$ is the
between-chain variance, and $W$ the within-chain variance,
\begin{align}
	B &= \frac{N}{M-1} \sum_{m=1}^M \left(\hat{\theta}_m - \hat{\theta} \right)^2\negthickspace, \\
	W &= \frac{1}{M} \sum_{m=1}^M \hat{\sigma}_m^2\negthickspace, 
\end{align}
where $\hat{\theta}$ is the parameter mean estimated using all samples, $\hat{\theta}_m$ is the mean 
estimated from a single chain, and $\hat{\sigma}_m^2$ is the parameter variance estimated from a 
single chain.

\section{Model comparison}

\hbindex{Bayesian model comparison} is based on \hbindex{\emph{model evidence}} (also known as \emph{marginal likelihood})
\begin{equation}
Z = \int P(\pvec)\;P(\dvec|\pvec)\;\ud\pvec,
\end{equation}
which we also recognise as the normalising constant in Eq.~\ref{eq:bayes_theorem}, i.e., the integrated posterior
density. Here, the multidimensional integration becomes the main problem. The integration can rarely be carried 
out analytically and numerical integration of multidimensional densities is far from trivial.

The use of model evidence in model comparison aims to penalise model complexity that is not justified by 
the data: since the posterior is the likelihood multiplied by the 
prior, and since each additional model parameter must be accompanied by a prior reflecting our knowledge
about the parameter,  the addition of a model parameter must increase the likelihood
more than the prior penalises the posterior (\citealt{Gregory2005}, Sect.~3.5; \citealt{MacKay2003}, Sect.~28.1).

Given two competing models, the \hbindex{posterior odds} in favour of model $M_1$ over model $M_2$ are
\begin{equation}
O_{12} = \frac{\pr(M_1|\dvec)}{\pr(M_2|\dvec)} = \frac{Z_1}{Z_2} \frac{P(M_1)}{P(M_2)}  = B_{12} \frac{P(M_1)}{P(M_2)},
\end{equation}
where $P(M_1)$ and $P(M_2)$ are the prior model probabilities and $B_{12}$ is called the \hbindex{Bayes factor}, 
the ratio of the model evidences. Given a set of models, the posterior model probabilities can be calculated from the posterior odds as
\begin{equation}
P(M_i|\dvec) = \frac{O_{i1}}{\sum_{1}^{n} O_{i1}},
\end{equation}
but model comparison is usually carried out based on odds or log-odds. If the models have equal prior probabilities,
the prior ratio equals to unity, and the model comparison can be done purely based on the Bayes factors.

\citet{Kass1995} provide general guidelines for interpreting Bayes factors or odds ratios, reproduced in 
Table~\ref{tbl:bayes_factors}. The limits are somewhat arbitrary, and should not be taken as hard rules.
Bayesian evidence can be very sensitive on the choice of priors, and sensitivity analysis should 
be carried out as a standard part of the analysis.

\begin{table}[t]
	\caption{Guidelines for interpreting Bayes factors, as presented by \citet{Kass1995}.}
	\label{tbl:bayes_factors}
	\begin{tabular*}{\columnwidth}{@{\extracolsep{\fill}} lll}
		\toprule\toprule
		$2\;\ln B_{10}$ & $B_{10}$ & Evidence against $H_0$\\
		\midrule
		0 to 2 & 1 to 3  &   Not worth more than a bare mention \\
		2 to 6 &  3 to 20 &   Positive \\
	    6 to 10 &  20 to 150 &  Strong \\
		$>$ 10 &  $> 150$ &  Very strong \\
		\bottomrule 
	\end{tabular*} 
\end{table}

Methods for evidence estimation start from simple approaches that can work with low-dimensional models,
such as direct Monte Carlo integration; to slightly more involved, such as different importance sampling
approaches; to complicated, such as the more advanced nested sampling approaches, bridge and path
sampling, and Bayesian quadrature. The methods are too numerous and intricate to be presented here, 
but are thoroughly reviewed by \citet{Clyde2007}, \citet{Ford2007}, and \citet[][Sect.~7]{Robert2007}.  

\hbindex{Nested sampling} \citep{Skilling2004,Skilling2006,Chopin2010} has gained popularity in astrophysics. The method 
provides both an evidence estimate and a sample from the posterior, and has publicly available implementations 
that can efficiently sample posteriors with multiple modes \citep{Feroz2009,Feroz2013a} and phase 
transitions \citep{Brewer}. The software packages are listed in Sect.~\ref{sec:tools}. 

Finally, \hbindex{Bayesian quadrature} \citep{Rasmussen2002,Osborne2012,Gunter2014,Hennig2015} offers an integration
approach that can be useful with relatively low-dimensional problems where the posterior evaluation is computationally 
expensive. Bayesian quadrature approximates the posterior density as a Gaussian process, which can be integrated
analytically, and allows for intelligent sampling.

As a final note, the Bayesian community is somewhat divided what comes to model comparison and the use of Bayesian 
evidence. Model comparison problems can often be transformed into parameter 
estimation problems, in which case a parameter estimation approach can advisable \citep[][Sect.~7.4]{Gelman2013}.

\section{Practical issues in exoplanet research}
\label{sec:practicalities}

While the theory behind Bayesian inference is straightforward, obtaining a reliable posterior estimate can be 
tricky in practise. MCMC samplers can fail to properly sample multimodal posteriors, and the samplers
can be inefficient if any of the parameters are strongly correlated. The optimal parametrisation for 
RV and transit  modelling, and how to deal with the stellar limb darkening in transit modelling, are especially
important practical issues in exoplanet research, and we briefly outline the current best practices below.

\subsection{Multimodal posteriors}

\hbindex{Multimodal posteriors} are rare in transit light curve modelling (indeed, a multimodal posterior in transit analysis
most often means that there is a problem in the code or the data), but quite common in radial velocity planet searches. 

The main issue with multimodal posteriors is that an MCMC sampler can fail to travel through the valleys between
the modes, and get stuck to sample the local mode nearest to where the chain was started from.  In
this case, the multimodality might not be visible from the posterior estimate at all, and we might end up with an overly
simplistic picture of the posterior space. This can be alleviated by 
\begin{itemize}
	\item starting multiple chains from random locations in the parameter space (for example, drawing the starting locations from the prior),
	\item starting multiple chains close to the posterior maxima obtained using a global optimiser,
	\item starting multiple chains close to the maxima estimated using local optimiser started from different points in the parameter space,
	\item or using an MCMC approach designed to sample multimodal distributions efficiently (such as parallel tempering MCMC).
\end{itemize}

\subsection{Parametrisation in transit and RV modelling}
\label{sec:practical_issues.parametrisation}

Basic MCMC algorithms, such as Metropolis-Hastings, are most efficient sampling approximately multivariate normal 
posteriors with minimal correlation between parameters (this is because the proposal distribution is usually 
multivariate normal). Inefficient sampling shows as a long autocorrelation time, and means that a long
chain is required to obtain a representative posterior sample. It generally makes sense to carry out the sampling 
with a parameter set that aims to minimise mutual correlations (sampling parametrisation), which is then mapped 
to the native model parametrisation.

Priors must be considered when deciding the \hbindex{sampling parameters}. A weakly informative (noninformative) prior
on a sampling parameter will not be a weakly informative prior on the model parameter if the mapping from
one to another involves nonlinear transformations \citep[][see the appendix]{Burke2007}. For example, a 
uniform prior on planet-star area ratio, $d = k^2$, where $k$ is the planet-star radius ratio, does not lead to 
a uniform prior on the radius ratio. While this is not a significant issue with parameters strongly constrained by the 
likelihood, it may lead to biases for parameters for which the observations do not yield a lot of information 
(that is, when the prior dominates the posterior). If one desires to set a weakly informative prior on a model
parameter that is mapped from the sampling parametrisation through a nonlinear transformation, one needs to
calculate the Jacobian of the transformation.

\citet{Ford2006,Carter2008,Kipping2010b} (among others) have investigated  how the sampling 
parameter set affects the efficiency of basic MCMC routines in the RV and transit light curve modelling. 
A transit model can be generally defined using from 7 to 11 parameters, and an RV model with five parameters.
Typical "physical" model parametrisations are
\begin{description}
	\item[Transit model:] 1) zero epoch; 2) orbital period; 3) orbital inclination; 4) semi-major axis; 5) eccentricity;
		6) argument of periastron; 7) planet-star radius ratio; and 8) $n$ limb darkening coefficients, depending on the limb darkening model
	\item[Radial velocity model:] 1) zero epoch; 2) orbital period; 3) eccentricity; 4) argument of periastron; and 5) $M \sin i$
\end{description}
but the optimal sampling parametrisation depends on the purpose of the analysis. Below we consider generic
transit or RV modelling, but, for example, RV planet searches can benefit from using $\log p$ as  a sampling parameter, 
where $p$ is the orbital period, instead of $p$.

\runinhead{Inclination and impact parameter}
In transit light curve modelling, orbital inclination $i$ can be substituted by the \hbindex{impact parameter} $b$. The mapping is
\begin{equation}
	b = a_s \cos i  \left[ \frac{1-e^2}{1+e\sin\omega} \right],  \qquad i = \arccos \left ( \frac{b}{a_s} \left[ \frac{1+e\sin\omega}{1-e^2} \right] \right ),
\end{equation}
where $a_s = a/R_\star$, that is, the semi-major axis divided by the stellar radius, $e$ is the eccentricity and $\omega$ 
the argument of periastron. The eccentricity-dependent term in square brackets equals to unity for circular orbits, and can be ignored
for small $e$. The impact parameter is bounded by $[0, 1+k]$ for a transiting planet, where $k$ is the planet-star 
radius ratio.

\runinhead{Semi-major axis and stellar density}
The scaled semi-major axis, $a_s$, can be replaced by transit duration \citep{Kipping2010b} or \hbindex{stellar density}, $\rho_\star$. The latter is 
practical since information from stellar characterisation can be used to create physically-based informative priors on 
stellar density, and the estimated stellar density can be used in planet candidate validation \citep{Seager2003,Tingley2011a,Kipping2013a}. 
The mapping can be derived from the Kepler's third law, and is
\begin{equation}
	a_s = \left( \frac{G \rho_\star p^2}{3\pi} \right)^{1/3}, \qquad \rho_\star= \frac{3\pi a_s^3}{Gp^2}
\end{equation}
where $G$ is the gravitational constant, $p$ is the period, and $\rho$ is the stellar density (all in SI units).

\runinhead{Eccentricity and argument of periastron}
Direct sampling in eccentricity, $e$, and argument of periastron, $\omega$, can be inefficient with most MCMC samplers.
The argument of periastron is not well constrained for low-eccentricity orbits, and it can have a bimodal
posterior if modelling only photometry without RV observations (such as when modelling light curves observed
over the whole orbital phase, where the phase curve and secondary eclipse yield information about $e$ and $\omega$).

\citet{Ford2005} suggested using $e\cos\omega$ and $e\sin\omega$ as an efficient sampling parametrisation, but 
noted later that setting an uniform prior on the parameters would set a non-uniform prior on the eccentricity,
leading to biased eccentricity estimates \citep{Ford2006}. \citet{Anderson2011d} improved the parametrisation
slightly into
\begin{equation}
a = \sqrt{e}\cos\omega, \qquad b = \sqrt{e}\sin\omega,
\end{equation}
which can be mapped to $e$ and $\omega$ as
\begin{equation}
e = a^2 + b^2, \qquad \omega = \arctan(b,a).
\end{equation} 
This parametrisation ensures that an uniform prior on $a$ and $b$ leads to an uniform prior on the 
eccentricity.

\subsection{Stellar limb darkening in transit modelling}
\label{sec:practical.limb_darkening}

Stellar \hbindex{limb darkening} (LD) is a significant source of uncertainty in transit modelling. The transit codes 
represent the stellar limb darkening profile as a linear combination of basis functions \citep{Mandel2002,Gimenez2006}.
The coefficients of these functions (\hbindex{limb darkening coefficients}, LDC) are mutually correlated and 
degenerate with the planet-star radius ratio and impact parameter. That is, a wide combination of 
LDC values, radius ratios and impact parameters may explain the observations equally well.

Several approaches have been used in attempt to overcome the degeneracies: the LDC can be fixed to 
theoretical values, they can be left completely unconstrained in order to marginalise over the whole
LDC space allowed by the data, or they can be constrained with informative priors based on theoretical
models.

\runinhead{Fixed limb darkening coefficients}
In the early phases of transiting exoplanet research, the limb darkening coefficients were usually fixed to values 
interpolated from limb darkening coefficient tables based on numerical stellar models, such as the tables by 
\citet{Claret2004,Claret2011,Claret2012,Claret2014} and \citet{Sing2010}. This is an easy way to circumvent
the complications due to limb darkening if the stellar models are reliable. However, if they are not (they
were not, see \citealt{Claret2008,Claret2009}), fixing the limb 
darkening coefficients leads to biased parameter estimates \citep[][but see also \citealt{Muller2013} 
for a counterargument]{Csizmadia2013,Espinoza2015}. 

The stellar models have improved during the last decade, and the current models can be considered more reliable
than the ones compared by \citet{Claret2008,Claret2009}. However, the possible biases are not the only problem
when fixing the LDCs. Even in the case of a perfect stellar model, we generally do not know the planet host star
perfectly. Fixing the LDCs does not allow us to account for the uncertainties in the stellar parameters, which leads
to underestimated uncertainties in the planet characterisation.

Thus, fixing the LDC cannot really be advocated in a Bayesian setting, since it can lead to 
biases and ignores the uncertainties in the stellar characterisation.

\runinhead{Unconstrained limb darkening coefficients}
A second approach is to leave limb darkening completely unconstrained, so that the radius ratio and impact parameter
estimates are marginalised over the whole limb darkening parameter space allowed by the data.
This is the best way to minimise the parameter estimate biases, but since the LD coefficients are mutually
correlated, the approach can reduce the MCMC sampling efficiency. 

Here again, the sampling efficiency can be improved by using a less correlated parametrisation. Using the quadratic
limb darkening model as an example, linear combinations of the two coefficients have often been used as sampling
parameters \citep[e.g.][]{Holman2006,Carter2009a}. Especially, a mapping proposed by \citet{Kipping2013b}
\begin{align}
u_1 = 2\sqrt{q_1} q_2, \qquad &u_2 = \sqrt{q_1} \left(1-2q_2\right),\\
q_1 = (u_1 + u_2)^2, \qquad &q_2 = \frac{u_1}{2(u_1 + u_2)},
\end{align}
where $u_1$ and $u_2$ are the quadratic coefficients, and $q_1$ and $q_2$ are the sampling parameters,
has proven practical. With most parametrisations, one needs to carry out separate tests to ensure that the
coefficients yield physically viable limb darkening profiles, but this mapping allows one to use uniform priors 
from 0 to 1 on $q_1$ and $q_2$ to cover the whole physically viable ($u_1, u_2$)-space.

\citet{Kipping2016} introduced a similar mapping for the three-parameter non-linear limb darkening model 
by \citet{Sing2009}, and provides code implementing the mapping, its inverse, and the analytical criteria
for the physical validity of the three-parameter LD model from \url{https://github.com/davidkipping/LDC3}.

\runinhead{Limb darkening coefficients constrained by model-based informative priors}
The third approach is a compromise between the two extremes. The LD coefficients can be assigned informative priors
based on stellar atmosphere models and the uncertainties in the host star characterisation. This approach allows one to 
use information from the models, but reduces biases and is less likely to seriously underestimate the parameter 
uncertainties. While the creation of priors is slightly more complicated than either of the two other approaches, 
tools exist to make the approach straightforward \citep{Parviainen2015,Espinoza2015}.

\section{Tools for Bayesian inference}
\label{sec:tools}

We finish the chapter with a list of some of the most matured tools for posterior estimation, nested sampling, and Gaussian 
processes that provide a \textsc{Python} interface. (We constrain the discussion to \textsc{Python} since it is currently the de facto
programming language for astrophysics. All major programming languages have similar packages available.) 

\runinhead{Posterior estimation} Several generic packages exist for posterior estimation using MCMC sampling.
\begin{description}[Type 1]
	\item[\textsc{PyMC3}] by  \citet{Salvatier2015} offers a versatile toolkit for Bayesian inference built on \textsc{Theano}. 
		The package includes a set of MCMC samplers, a Gaussian-process framework, and tools
		for working with more advanced topics, such as mixture models and variational inference.
		The package is available from \url{https://github.com/pymc-devs/pymc3}, and includes extensive documentation 
		with tutorials and examples as IPython notebooks, which makes it very useful for learning Bayesian inference in practise. 
	\item[\textsc{Stan}] (\url{http://mc-stan.org}) is a probabilistic programming language that offers similar functionality as 
		\textsc{PyMC3}, but with interfaces for many programming languages. For example, \textsc{PyStan}  (\url{https://github.com/stan-dev/pystan}) 
		provides an interface for \textsc{Python},  and \textsc{RStan} (\url{https://github.com/stan-dev/rstan})  for \textsc{R}. 
	\item[\textsc{Edward}]  is a Python library for probabilistic modelling  \citep{Tran2016} built on \textsc{TensorFlow}, a machine learning framework
		developed by Google. \textsc{Edward} fuses three fields: Bayesian statistics and machine learning, deep learning, and probabilistic programming.
		\url{http://edwardlib.org}
	\item[\textsc{emcee}] by \citet[][ \url{https://github.com/dfm/emcee}]{Foreman-Mackey2012} is a popular MCMC sampler
		implementing the affine-invariant ensemble sampler by \citet{Goodman2010}. The sampler is efficient in sampling
		correlated parameter spaces, and does not require the ability to calculate posterior derivatives, unlike many of the advanced 
		MCMC samplers. The implementation also offers straightforward MPI parallelisation, which  allows the sampling
		to be carried out efficiently in a cluster environment.
\end{description} 

\runinhead{Nested sampling} Of all the evidence estimation methods, nested sampling is currently the most
popular (and accessible) one used in astrophysics. Nested sampling covers a variety of sampling methods, 
each with their advantages and disadvantages, and the research of new approaches is vibrant. 
\begin{description}[Type 1]
	\item[\textsc{MultiNest}] implements a nested sampling algorithm written in \textsc{Fortran 90} that excels in 
	sampling multimodal densities with strong degeneracies  \citep{Feroz2009,Feroz2013a}. \textsc{MultiNest} 
	draws samples from a constrained prior distribution approximated with ellipsoids, which makes it efficient
	for moderate dimensional problems (up to $\sim$40), but other approaches can work better for high-dimensional
	problems \citep{Brewer2016}. The code with \textsc{C/C++} wrappers is available from 
	\url{www.mrao.cam.ac.uk/software/multinest}, and a \textsc{Python} wrapper, \textsc{PyMultiNest}, is available
	from \url{https://github.com/JohannesBuchner/PyMultiNest}.
	\item[\textsc{DNest}] aims to sample multimodal distributions with degeneracies between parameters efficiently, like
	\textsc{MultiNest}, but uses MCMC for sampling \citep{Brewer,Brewer2016}. The code is written in \textsc{C++}, 
	supports \textsc{Python} and \textsc{Julia}, and is available from \url{https://github.com/eggplantbren/DNest4}. 
	\item[\textsc{Nestle}] provides a pure-\textsc{Python} implementation of nested sampling using either MCMC sampling,
	single-ellipsoid sampling, or \textsc{MultiNest}-like multi-ellipsoid sampling. The code is  available from \url{https://github.com/kbarbary/nestle}.
\end{description}

\runinhead{Gaussian processes} In their most basic form, Gaussian processes (GPs) can be implemented with 3-5 lines of \textsc{Python} code using
\textsc{SciPy}'s linear algebra module. However, more advanced functionality--such as the ability to construct complex kernels, combine
kernels, and solve the GP with better than $\mathcal{O}(n^3)$ scaling--requires more involved approaches.
Gaussian processes are widely used by the machine learning community, and are usually implemented in one form or another in
various machine learning packages.
\begin{description}[Type 1]
	\item[\textsc{George}] is a Gaussian process framework that implements two GP solvers: a basic solver that uses \textsc{SciPy}'s 
		Cholesky decomposition, and a HODLR solver implementing the \citet{Ambikasaran2014} approach 
		that scales as $\mathcal{O}(n \log^2 n)$. The package has been used widely by the astrophysics community, and is available from 
		\url{https://github.com/dfm/george}.
	\item[\textsc{Celerite}]  implements the $\mathcal{O}(n)$ GP solver described by \citet{Foreman-Mackey2017,Ambikasaran2015}. The approach
		is restricted to 1D processes where the covariance kernel is represented as a sum of complex exponentials, but is flexible enough to model a wide variety
		of astrophysical variability. The code with \textsc{C++} and \textsc{Python} interfaces is available from \url{https://github.com/dfm/celerite},
		and a \textsc{Julia} implementation can be found from \url{https://github.com/ericagol/celerite.jl}. 
	\item[\textsc{GeePea}] is a general GP package developed by Neale Gibson \citep[][\url{https://github.com/nealegibson/GeePea} ]{Gibson2011a}. 
		The package is lightweight and easily extensible, and has been widely used in exoplanet transmission spectroscopy.
	\item[\textsc{GPy}] is a Python GP framework developed by the Sheffield machine learning group. The package offers a wide variety of
		kernels, and is available from \url{https://github.com/SheffieldML/GPy}.
	\item[\textsc{GPFlow}] is a GP package using \textsc{TensorFlow} \citep{Matthews2016}. The package has been developed following \textsc{GPy},
	but uses the \textsc{TensorFlow} machine learning framework to allow for faster and bigger computations. The package can be found from \url{https://github.com/GPflow/GPflow}.
	\item[\textsc{PyMC3}] offers a lightweight but versatile GP framework with a set of basic covariance functions, and provides tools to 
	modify the covariance matrix directly.
	\item[\textsc{scikit-learn}] is a multi-purpose machine learning package that includes Gaussian processes. The package is available from 
		 \url{https://github.com/scikit-learn/scikit-learn}.
\end{description}

\section{Literature and resources}
\label{sec:literature}

Bayesian inference is a wide-ranging topic with books devoted to various subfields. This chapter has scratched the surface in the hope this will allow
the reader to understand the basic concepts commonly encountered in Exoplanet research, but has not tried to delve into any of the 
more advanced topics, such as hierarchical modelling. However, in order to apply Bayesian methods in research, one should
read at least one of the entry-level books dedicated to Bayesian inference
\begin{itemize}
	\item \citet{Gregory2005} covers the basics of Bayesian inference in an astrophysical context, with several exoplanet-related examples.
	\item \citet{Gelman2013} offers an overarching treatise to Bayesian inference, starting from the basics, but also covering a wide variety of advanced topics.
	\item \citet{Robert2007} offers an in-depth look into Bayesian inference from the perspective of decision theory.
	\item \citet{MacKay2003} covers Bayesian inference (amongst other topics) from the perspective of information theory.
\end{itemize}

Numerical methods improve continuously, and the landscape of generally accessible computational tools changes
quickly. Internet is an important source of up-to-date information. Websites, such as Cross validated \url{https://stats.stackexchange.com} 
(choose tag "bayesian"), offer a way to connect researches not specialised to statistics with specialists,
and statistics blogs, such as \url{http://andrewgelman.com} and \url{https://xianblog.wordpress.com}, 
can help to keep one up-to-date with current developments in the field.

\bibliographystyle{spbasicHBexo}
\bibliography{parviainen_exoplanet_handbook}
\end{document}